\documentclass{ws-procs9x6}            

\begin{document}
\title{Total kinetic energy release in the fast neutron induced fission of $^{232}$Th and $^{235}$U}

\author{W. Loveland$^*$ and J. King}

\address{Chemistry Department,Oregon State University\\
Corvallis, OR 97331, USA\\
$^*$E-mail: lovelanw@onid.orst.edu}

\begin{abstract}
We have measured the total kinetic energy release (TKE), its variance and associated fission product distributions for the 
neutron induced fission of $^{232}$Th and $^{235}$U for E$_{n}$ = 2-90 MeV.  The neutron energies were determined on an 
event by event basis by time of flight measurements with the white spectrum neutron beam from LANSCE.  The TKE decreases non-linearly 
with increasing neutron energy for both systems, while the TKE variances are sensitive indicators of n$^{th}$ chance fission.  The 
associated fission product distributions show the decrease in TKE with increasing beam energy that is due to the increasing probability of 
symmetric fission --which has a lower associated TKE-- and the decreasing TKE associated with asymmetric fission, presumably due to the  
decreasing importance of the A $\sim$ 132 shell structures. 
\end{abstract}

\keywords{Fission; TKE; variance of TKE; fission product mass distributions; $^{232}$Th(n,f); $^{235}$U(n,f)}

\bodymatter

\section{Introduction}
We report the total kinetic energy release (TKE) , the variance ($\sigma$$^{2}$) of the TKE distribution, and the associated mass yield distributions in the fast neutron induced fission of $^{232}$Th and $^{235}$U(E$_{n}$ = 2-90 MeV).  (A preliminary version of the $^{235}$U study has been reported before \cite{yanez}.)  While these 
data have implications for nuclear weapons and reactors, our primary goal is to better understand the fission process.  As we increase the energy input to these 
systems from $\sim$ 0 to $\sim$ 90 MeV, we find that the primary energy released in fission, the TKE, decreases by a few MeV (6.0 and 7.4 MeV, for $^{232}$Th and 
$^{235}$U, respectively).  Clearly the dissipation of the incident neutron energy in the fissioning system takes place on a time scale ($\sim$ 10$^{-21}$ s) that is 
short compared to the times for the large scale collective motion of fission ($\sim$ 10$^{-18}$ s).  The dissipated energy appears in the excitation energy of the 
fission fragments while the small decrease in the observed TKE reflects changes in the fission process in these reactions.  Understanding and characterizing  these processes will be the central theme of this work.  By studying the $^{232}$Th(n,f) and  the $^{235}$U(n,f) processes simultaneously, we hope to compare and contrast the physics of the different fissioning systems upon the fission product mass and energy distributions.

\section{Experimental }
This experiment was carried out at the Weapons Neutron
Research Facility (WNR) at the Los Alamos Neutron Science
Center (LANSCE) at the Los Alamos National Laboratory \cite{lansce1,lansce2} over a 5 day period in December, 2014.   ``White spectrum" neutron beams were generated from
an unmoderated tungsten spallation source using the 800 MeV
proton beam from the LANSCE linear accelerator. The measured white spectrum near
the target position is shown in Fig. 1. The proton beam
is pulsed allowing one to measure the time of flight (energy)
of the neutrons arriving at the experimental area. Note the beam intensities of $\sim$ 10$^{5}$ n/s --a number that is higher than the beam used at 
radioactive beam facilities--but substantially less than that normally used in fission studies.  Thus one has to run for long periods of time and/or to use large 
solid angle detection systems.

\begin{figure}[ht]
\includegraphics[scale=0.3]{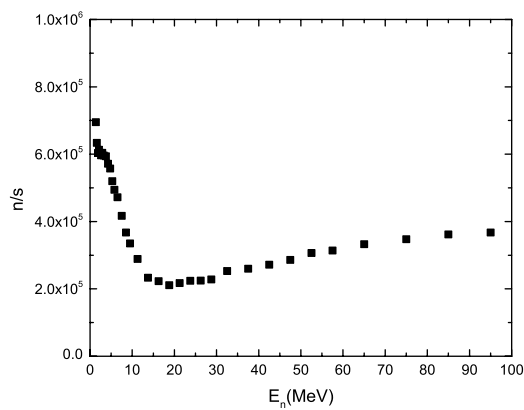}
\caption{Measured neutron beam intensity near  the target location on the 15R beam line during this experiment}
\label{aba:fig1}
\end{figure}

The detailed description of this experiment has been reported previously \cite{yanez}.  We refer the reader to \cite{yanez} for details.
 In \cite{yanez}, assuming $\nu$$_{tot}$ = 2.43 \cite{ref24} and $\nu$$_{pre}$ = 0.0 \cite{ref24}, we found that the measured thermal neutron induced post fission TKE was E$_{TKE}$$^{th. n}$= 169.8 $\pm$ 0.4 MeV and that $\sigma$$_{TKE-th}$$^{2}$=103.6 $\pm$ 0.2 MeV.  The pre-fission TKE is E$_{TKE}$$^{*,th}$= 170.7 $\pm$ 0.4 MeV.  This result is consistent with the previous measurements of 171.9 $\pm$ 1.4 \cite{ref22}, 172.0 $\pm$ 2.0  \cite{ref25} and 170.1 $\pm$ 0.5 \cite{bm}. 
For the spontaneous fission of $^{252}$Cf, we found the average post fission light fragment TKE was 103.2 $\pm$ 1.1 MeV while the average heavy fragment TKE was 78.6 $\pm$ 1.5 MeV.  This is to be compared to post neutron emission values \cite{Hal} for the light fragment of 103.8 $\pm$ 0.5 MeV and heavy fragment energy of 79.4 $\pm$ 0.4 MeV. {\bf Our measurements of the TKE should be regarded as absolute measurements.}

\section{Results and Discussion}
The TKE data for each system was binned into neutron energy bins.  The resulting TKE distributions for each neutron energy bin were found to be Gaussian in shape with no observed deviations from a Gaussian shape.  The mean values of the TKE for each bin for each system are shown in Figure 2.  The mean TKE values decrease non-linearly with increasing neutron energy in a manner that can be represented by a polynomial in neutron energy.  These data are in good agreement with previous measurements at lower energies \cite{madland, duke, trochon, sergachev, kiesewetter}.  The Viola systematics \cite{vic}, which applies only to pre-neutron emission data, would predict TKE values of 170.1 and 163.8 MeV, respectively, for the $^{235}$U(n,f) and $^{232}$Th(n,f) reactions.  That ``Viola ratio" (170.1/163.8) is remarkably similar to the observed TKE ratio for the two systems, perhaps indicating the same underlying physics.  This observation of a non-linear decrease in TKE with increasing excitation energy is certainly not unique--as it was observed in $^{235}$U(p,f) reactions decades ago by Ferguson et al. \cite{ferggy}.  Ferguson et al.  attributed this decrease to an increase in symmetric fission (which has a lower TKE) and the possible ``washing out" of the A=132 shell structure with increasing beam energy although they could not resolve the two contributing effects.  Using a modal analysis, Brosa et al.  \cite{brosa1} attempted to resolve this issue for these systems ($^{232}$Th(n,f) and $^{235}$U(n,f)) for neutron energies of E$_{n}$ = 0 - 6 MeV.  They concluded that the reduced TKE as a function of E$_{n}$ could be attributed simply to a increase in symmetric fission.  A similar result using an ``empirical fission potential" was achieved by  by Wang et al.\cite{wang}.  The choice of this low neutron energy regime is not accidental in that it assures that only single chance fission is being studied.

The variances of the TKE distributions for the $^{235}$U reaction are shown in Figure 3 along with the unpublished data of Duke \cite{duke}.  It is striking to note the occurrence of peaks in the variances at the same neutron energies in both works and the correspondence of those peaks with the onset of n$^{th}$ chance fission.  The variances of the TKE distributions represent a sensitive tool for detecting the occurrence of multiple chance fission.

\begin{figure}[ht]
\includegraphics[scale=0.3]{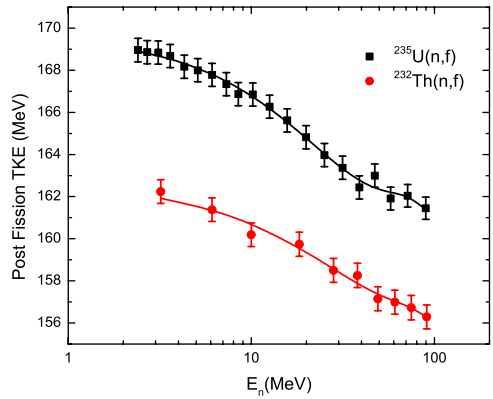}
\caption{Measured values of the mean TKE as a function of neutron energy for $^{232}$Th (n,f) and $^{235}$U(n,f).The lines are polynomial fits to the data, TKE(235) = 169.68-0.33881E$_{n}$-0.0052E$_{n}$$^{2}$-2.73475x10$^{-5}$ E$_{n}$$^{3}$ and TKE(232)=162.56-0.20877E$_{n}$+0.00275E$_{n}$$^{2}$-1.3367x10$^{-5}$ E$_{n}$$^{3}$.}
\label{aba:fig2}
\end{figure}

\begin{figure}[ht]
\includegraphics[scale=0.3]{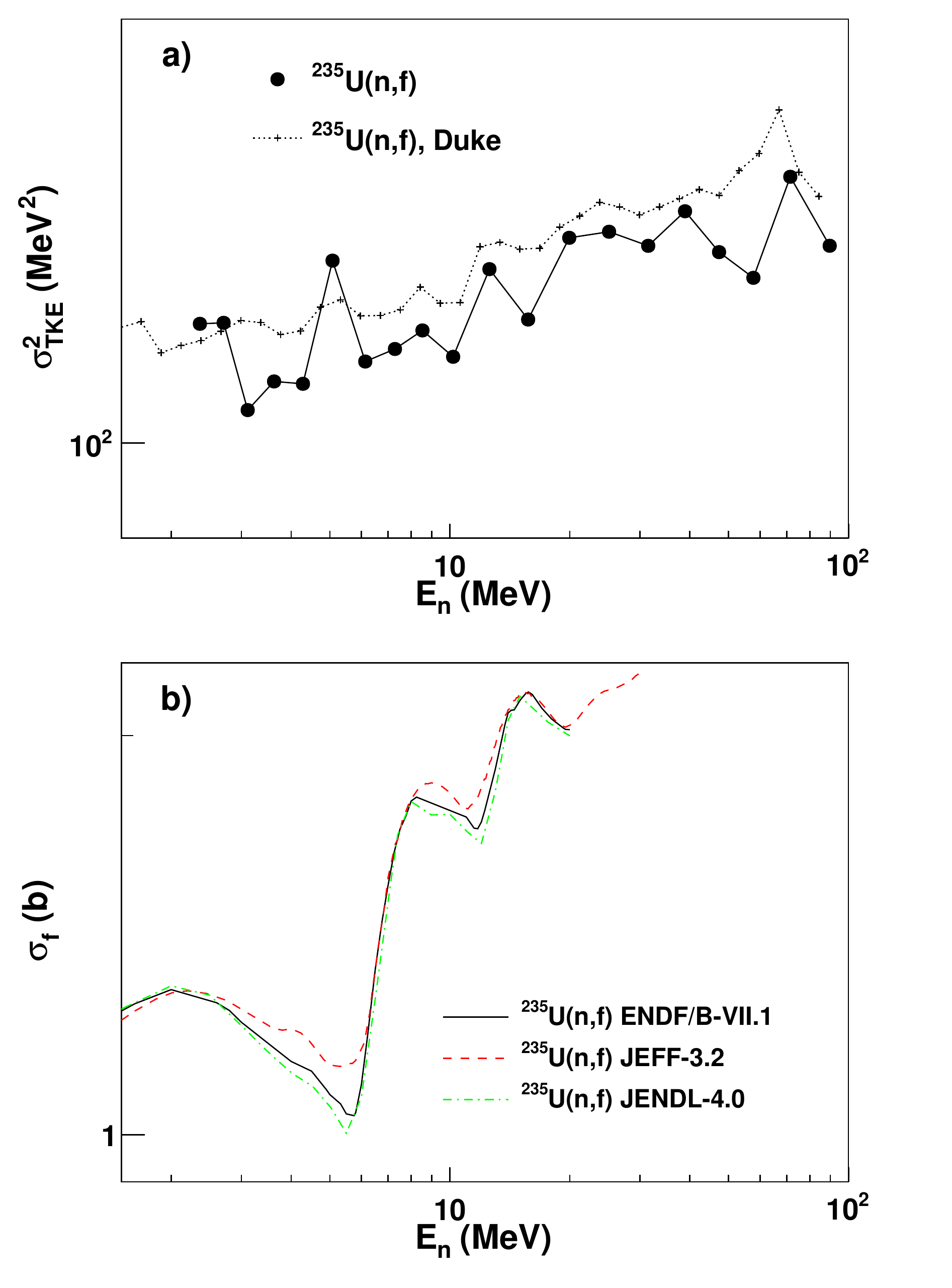}
\caption{(a) The measured variances of the TKE distributions as a function of neutron energy for the $^{235}$U(n,f) reaction.  (b) The fission cross section as a function of neutron energy for the  $^{235}$U(n,f) reaction.  }
\label{aba:fig3}
\end{figure}

The pre-neutron emission fission product mass yield distributions for the two reactions are shown in Figures 4 and 5.  Uncertainties in the individual points cam be deduced from the Y(A) values which are the number of counts in a given mass bin.  For both systems, one observes increasing amounts of symmetric fission with increasing neutron energy.  The $^{235}$U distributions contain about 75K events while the $^{232}$Th distributions contain about 20 K events so the uncertainties are smaller in the $^{235}$U data, but because of the shapes of the distributions, the symmetric component is more noticeable in the $^{232}$Th data.

In Figure 6(a), we show the evolution with neutron energy of the mass sorted TKE values for the $^{232}$Th(n,f) reaction.  One sees the TKE associated with symmetric fission is roughly constant with increasing E$_{n}$ with a lower average TKE value.  The TKE associated with asymmetric fission decreases with increasing E$_{n}$ which is incompatible with the notion that overall TKE decrease with increasing E$_{n}$ is due {\bf solely} to the increased probability of symmetric fission. Further confirmation of this trend is seen in Figure 6 (b) where one can observe the steep changes in the TKE associated with the A $\sim$ 134 fragments (associated with the spherical shell structures at Z=50, N=82, which accompany the trend of the symmetric fission TKE with increasing E$_{n}$.  

\begin{figure}[h]
\includegraphics[scale=0.5]{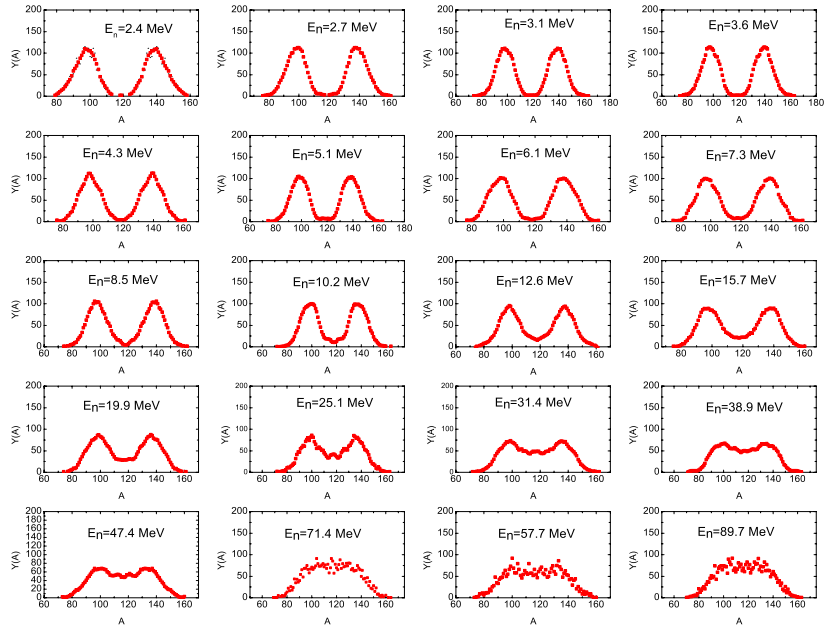}
\caption{The measured pre-neutron emission fission product mass distributions  as a function of neutron energy for the $^{235}$U(n,f) reaction.   }
\label{aba:fig4}
\end{figure}

\begin{figure}
\includegraphics[scale=0.5]{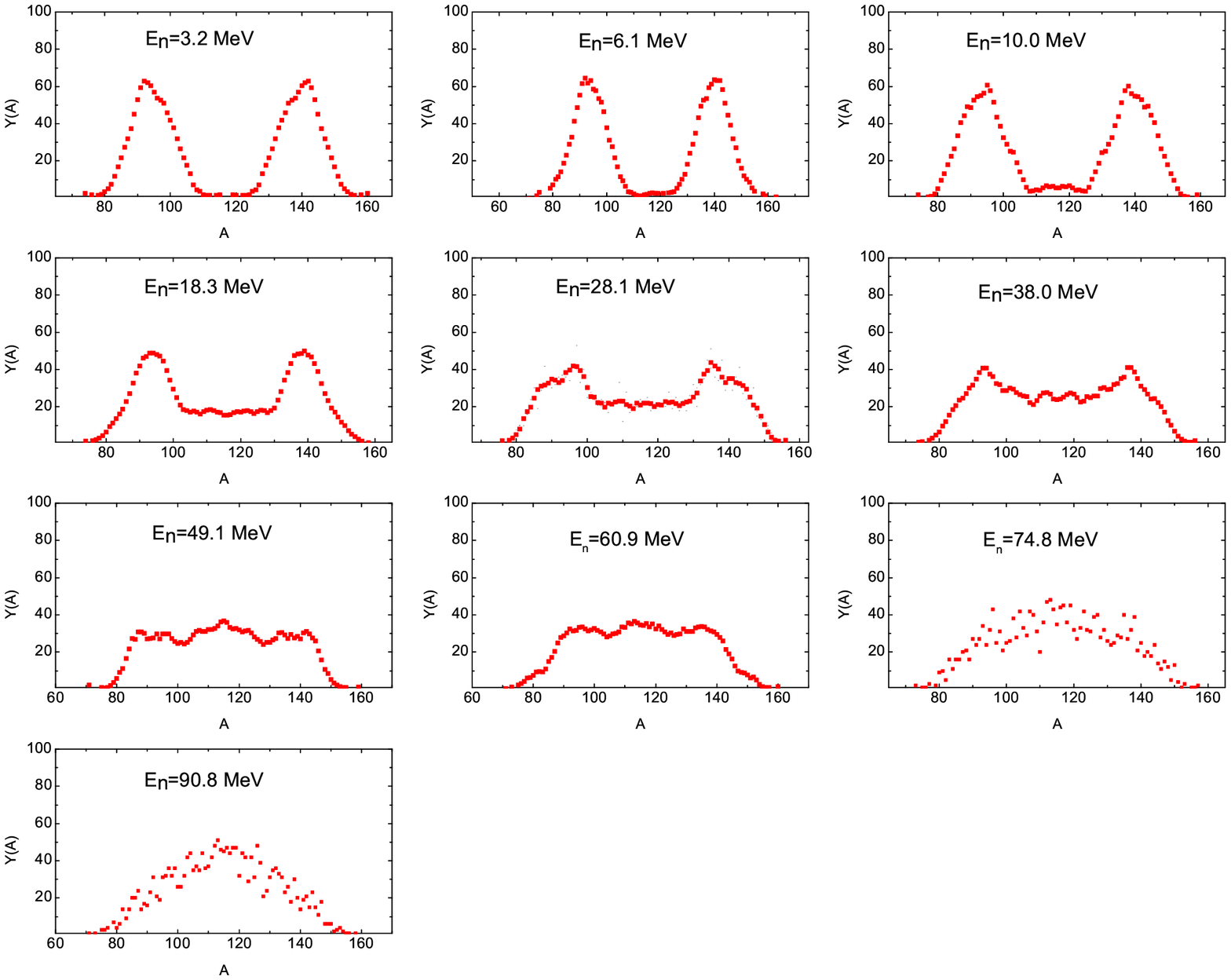}
\caption{(a) The measured pre-neutron emission fission product mass distributions  as a function of neutron energy for the $^{232}$Th(n,f) reaction. }
\label{aba:fig5}
\end{figure}

\begin{figure}[ht]
\includegraphics[scale=0.4]{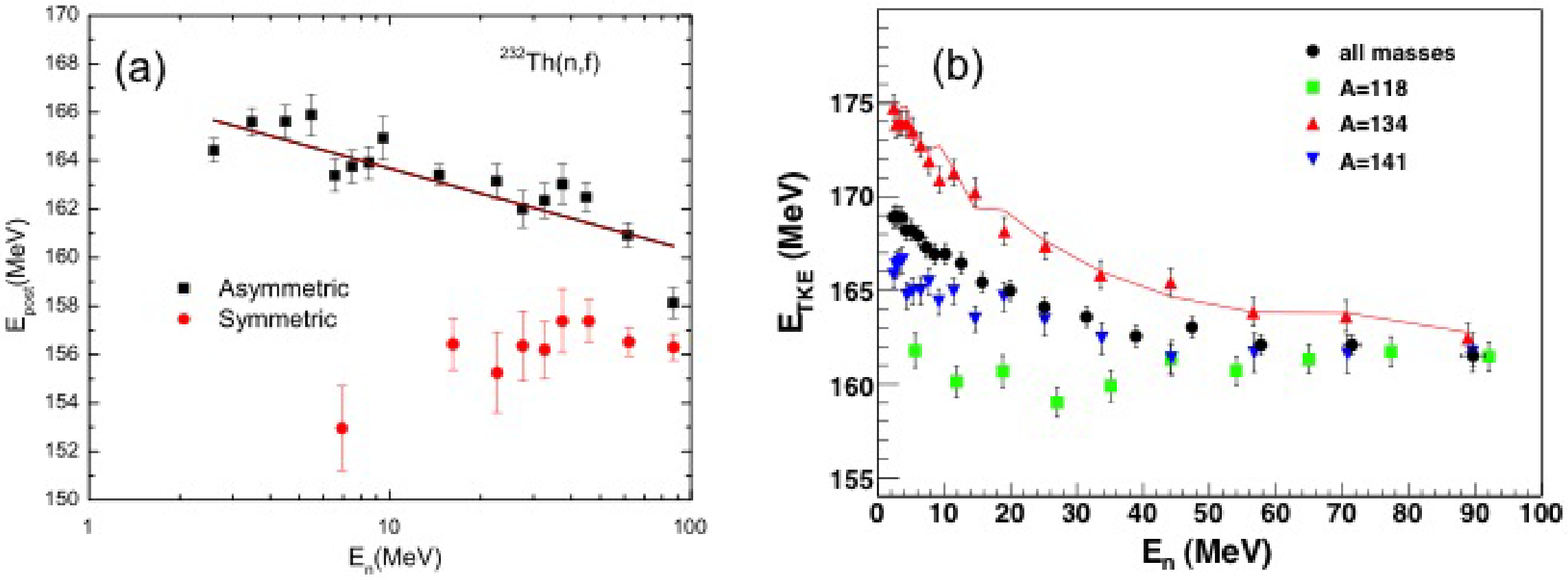}
\caption{(a) The mass sorted TKE values as a function of neutron energy for $^{232}$Th(n,f).  (b) The mass sorted TKE values as a function of neutron energy for $^{235}$U(n,f).}
\label{aba:fig6}
\end{figure}

To learn about fission, one would like to compare these measured properties of the TKE distributions with predictions of nuclear theory.  The theories should give ``predictive" estimates of measured quantities rather than ``post-dictive" estimates.   Two semi-empirical models \cite{GEF,Lestone} for the TKE release as a function of neutron energy are compared to our data in Figure 7.  Both models take into account multiple chance fission.  The predictions of the GEF code differ from the data above E$_{n}$ = 20 MeV while the predictions of \cite{Lestone} agree quite well with the data.  One should note that more ``fundamental" calculations of fission product yields \cite{xx} do exist for these systems.  In these calculations, there is a washing out of the shell effect and changes in the pairing energies with increasing excitation energy.  The calculation \cite{xx} is limited to predicting (correctly) the fission product mass distribution for $^{235}$U(n,f) at E$_{n}$=14.5 MeV.

\begin{figure}[ht]
\includegraphics[scale=0.4]{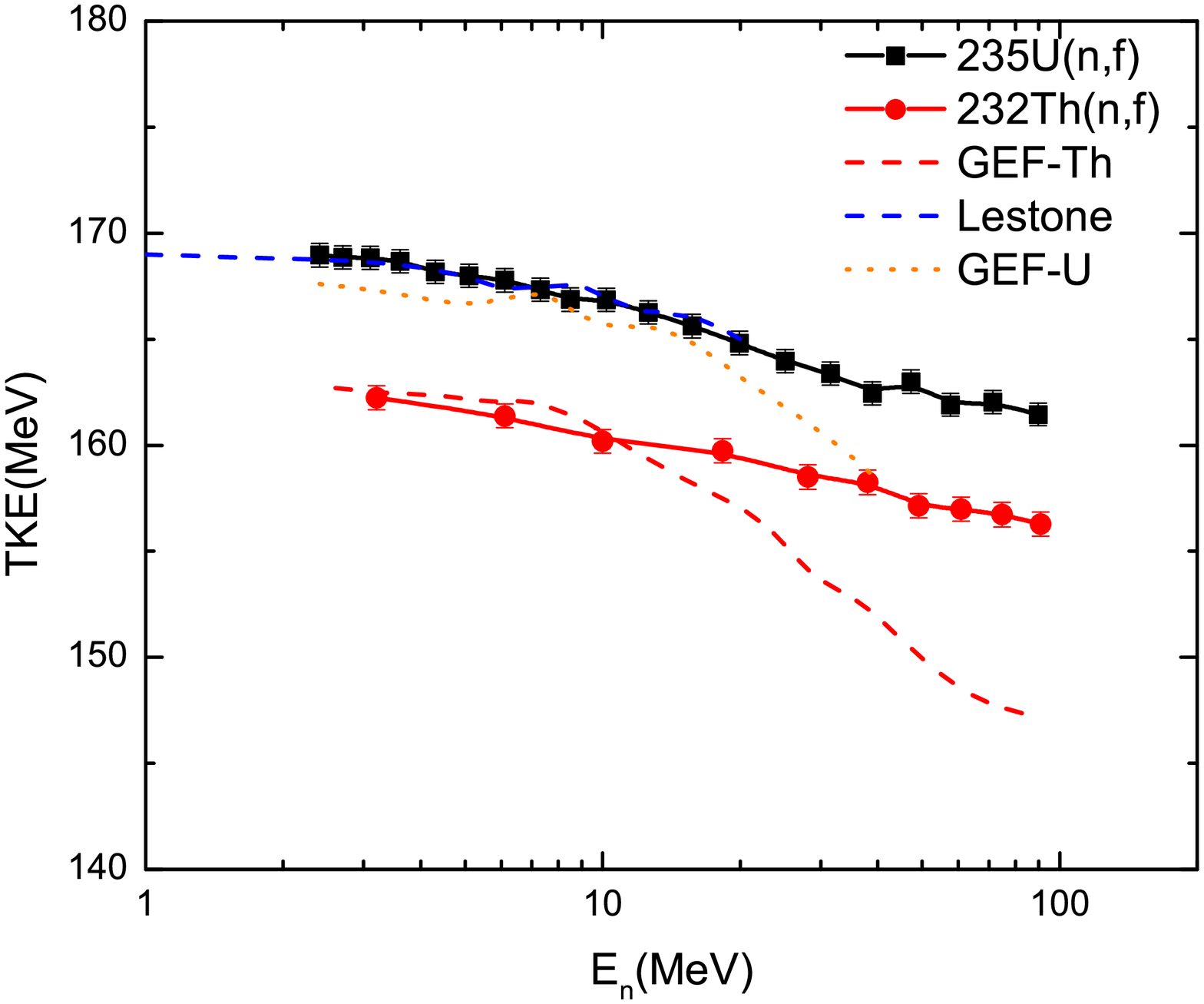}
\caption{Comparison of our measured TKE values  with the predictions of \cite{Lestone} and \cite{GEF}}
\label{aba:fig7}
\end{figure}

\section{Acknowledgments}
This work was supported in part by the U.S. Dept. of Energy, NNSA, under Grant DE-NA0002926 and the U.S. Dept. of Energy, Office of Science, Office of Nuclear Physics under award number DE-SC0014380. We gratefully acknowledge the contributions of Ricardo Yanez to this work.

\end{document}